# Numerical analysis of effective refractive index ultrasonic sensor based on Cantilever arm structure slot-based dual-micro-ring resonator

C. Y. Zhao,[a,b]* P. Y. Li[a], C. M. Zhang[c]

[a]College of Science, Hangzhou Dianzi University, Zhejiang 310018, China

[b]State Key Laboratory of Quantum Optics and Quantum Optics Devices, Shanxi University, Taiyuan 030006, China

[c]Nokia Solutions and Networks, Hangzhou, 310053, China

**Abstract:** We propose a novel ultrasonic sensor structure composed of Cantilever arm structure slot dual-micro-ring resonators (DMRR). We present a theoretical analysis of transmission by using the coupled mode theory. The mode field distributions and sound pressure distributions of transmission spectrum are obtained from 3D simulations based on Comsol Multi-physics (COMSOL) method. Our ultrasonic sensor exhibit theoretical sensitivity as high as $1462.5 mV/kPa$, which is 22 times higher than that of single slot-based micro-ring ultrasonic sensor. Our ultrasonic sensor offers higher sensitivity and a larger detection frequency range than conventional piezoelectric based ultrasound transducer. The results show that the sensing characteristics of our system can be optimized through changing the position and the angle of sound field. Our ultrasonic sensor with an area of $25 \mu m \times 60 \mu m$, the $Q$-factor can be approximately $1.54 \times 10^3$ with radius of $5 \mu m$. We detect an angular range of -90º to 90º and a minimum distance of $0.01 \mu m$. Finally, we calculate the Cantilever arm structure slot DMRR array ultrasonic sensor's optical performance. Our proposed design provides a promising candidate for a hydrophone.

Corresponding author. Email: zchy49@hdu.edu.cn.

**1.Introduction**

A hydrophone can convert acoustic signal into electrical signal, which has many important applications in underwater communication, detection, target positioning and tracking. Traditional hydrophone is mostly made of piezoelectric ceramic, the mechanical stress put on piezoelectric ceramic will cause the piezoelectric effect. In 1983, N. Yanagihara firstly designed a vibrator receive the sound signal based on piezoelectric effect.[1] N. Felix *et al*. found that the electro-acoustic characteristics of the two-dimensional transducer array were superior to that of the traditional piezoelectric ceramics.[2] G. J. Zhang *et al*. designed a MEMS hydrophone in order to obtain a more accurate sound signal.[3] X. B. Yang verified that Cylindrical transducer array can better control the focus of sound field.[4] J. S. Wang designed a PMN-PH-PT piezoelectric ceramics high-temperature transducer.[5] A. Tikhomirov *et al*. studied the DFB fiber hydrophone array in order to reduce the cost of materials.[6,7] Considering the traditional piezoelectric ceramic volume is large, S. Y. Xu put forward to a compact optical fiber hydrophone,[8] H. Li *et al*. proposed a Cylindrical optical fiber acoustic sensor.[9] The external stress will cause thermal fatigue, so the heat loss will increase, K. Saijyou *et al*. designed a novel optical fiber hydrophone in order to reduce Sinusoidal noise.[10] Marcatill proposed the concept of micro-ring resonator(MRR) in the 1970s.[11] With the manufacturing process of micro-nano photonic devices become more and more mature,[12-17] MRRs are also widely used in the design of hydrophones due to low loss, low cost, and ease to produce.[18,19] Firstly, the liquid environment can change the refractive index of MRR cladding region. Secondly, the material itself will cause water absorption and water loss. Thirdly, the propagation, absorption and loss of optical signal will change the refractive index of medium, the ultrasonic wave has the following characteristics: less energy consumption, a long distance transmission in the liquid environment. The elastic-optical effect of ultrasonic wave can changes the effective refractive index of the system, which furthermore results in the resonance peak shifts. C.Y. Chao *et al.* firstly designed a MRR ultrasonic sensor with a frequency up to $50 MHz$.[20] L. Tao *et al.* produced a polymer MRR ultrasonic sensor [21] and R. Rekha *et al.* proposed a Cantilever arm type MRR sensor.[22] C. Zhang *et al.* realized the polymer MRR hydrophone with a frequency up to 350 MHz.[23]

In 2004, M. Lipson firstly proposed the slot wave-guide structure.[24] C. A. Barrios *et al.* produced a slot-based MRR bio-sensor with a sensitivity up to $212 nm/RIU$ to realize a label-free molecular detection.[25-26] J. T. Robinson *et al.* made a slot-based MRR cavity with a diameter of $20\mu m$, the sensitivity reaches to $490 nm/RIU$.[27] S. L. He *et al.* proposed a slot-based MRR ultrasonic sensor with a radius of $12\mu m$, the sensitivity is $66.7 mV/kPa$, and the frequency up to $540 MHz$.[28] C. M. Zhang *et al.* used the finite-different time-domain (FDTD) method to simulate and calculate the optical and acoustic characteristics of a slot-based MRR acoustic sensor with a radius of $5\mu m$, the sensitivity is $2453.7 mV/kPa$.[29] Based on the above analysis, in this paper, we adopt a Cantilever arm structure slot-based DMRRs cavity to investigate the ultrasonic sensing characteristics. Then, we use the FDTD simulation software and COMSOL simulation software to investigate the influence of the distance and the radiation angle of the sound field on the ultrasonic sensing characteristics.[30]

## 2. Theoretical Model of Slot-based DMRRs

The structural diagram of GaN straight wave-guide and DMRRs slot wave-guides as shown in Fig.1(a). The cross-section of the slot structure as shown in Fig.1(b).

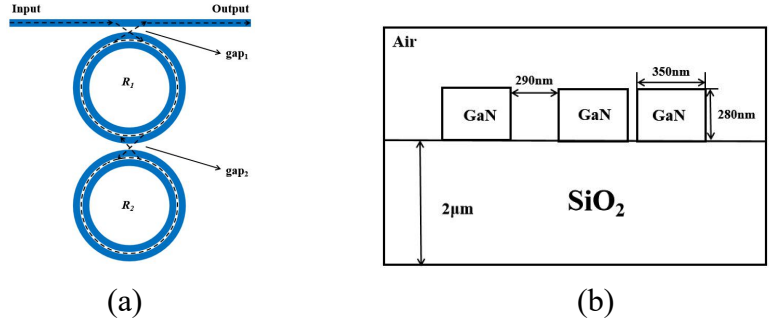

**Fig.1.** (a)The structure of a slot-based DMRRs, (b)the cross-section diagram of slot structure.

The incident beam is injected into the input port, the Electromagnetic field can be expressed as a sum of the wave-guide TE modes, and finally transmit them through the output port(see the dotted line in Fig1(a)). The system was fabricated on commercial SOI wafers, the system parameters are: the straight wave-guide length of $25\mu m$, the outer ring radius of $5\mu m$, the slot width is $75nm$, the $gap_1$ between the straight wave-guide and the first ring wave-guide is $290nm$, the $gap_2$ between the two rings wave-guides is also $290nm$. The wave-guide width of $350nm$ and the wave-guide height of $280nm$.

When the coupled light has a phase shift of an integer multiple of $2\pi$ along the DMRRs, according to the transfer matrix method, the relationship between output and input ports can be write as [31]

$$\begin{bmatrix} E_{out1} \\ E_{out2} \end{bmatrix} = \begin{bmatrix} t & -jk \\ -jk & t \end{bmatrix} \begin{bmatrix} E_{in1} \\ E_{in2} \end{bmatrix}, \quad (1)$$

The transmission of DMRRs is[31]

$$T = 1 - \frac{(1-r_2^2)(1-a_2^2|\tau_1|^2)}{(1-r_2 a_2 |\tau_1|^2)((1+4r_2 a_2 |\tau_1|)/(1-r_2 a_2 |\tau_1|^2))\sin^2[(\phi_{(eff)}+\phi)/2]}, \quad (2)$$

For two rings: $r_1$ and $r_2$ are transmission coefficients, $\alpha_1 = 0.7\,dB/cm$ and $\alpha_2 = 0.66\,dB/cm$ are the transmission losses coefficients, $L = 2\pi R$ is the perimeter when the radius $R = 5\mu m$, $a_1 = e^{-\alpha_1 L/2}$ and $a_2 = e^{-\alpha_2 L/2}$ are the attenuation factors, $\phi = 2\pi/\lambda n_{eff} L$ is the phase, $n_{eff}$ is the effective refractive index, $\lambda$ is the wavelength of incident light, $\tau_1 = (r_1 - a_1 e^{i\phi})/(1 - r_1 a_1 e^{i\phi})$ is the change of light intensity in the first ring, $\tau_2 = (r_2 - a_2 e^{i\phi})/(1 - r_2 a_2 e^{i\phi})$ is the change of light intensity in the second ring, $\phi_{(eff)} = \pi + \phi + \arg(\tau_1)$ is the equivalent phase.

The sound pressure can produce the material deformation[28]

$$\varepsilon = -\frac{(1+\upsilon)(1-2\upsilon)P}{E}, \quad (3)$$

where $\upsilon$ and $E$ are Poisson's ratio and Young's modulus of the material, respectively, and $P$ is the ultrasonic pressure. We take $\upsilon = 0.252$, $E = 4.611\,GPa$ and $P = 1MPa$.

A elastic-optical effect of cladding material in the $x$ direction can be expressed as[32]

$$n_x = n_0 - C_1 \sigma_x - C_2(\sigma_y - \sigma_z), \quad (4)$$

where $n_0$ is the refractive index of the material, we make $n_0 = 1.53$. $n_x$ means in the $x$ direction, $C_1$ and $C_2$ are the elastic modulus of the material and $\sigma_{x,y,z}$ is the sound pressure in different directions.

The comparison of transmission spectrum between the numerical calculation (a) and FDTD simulation (b) as shown in Fig. 2.

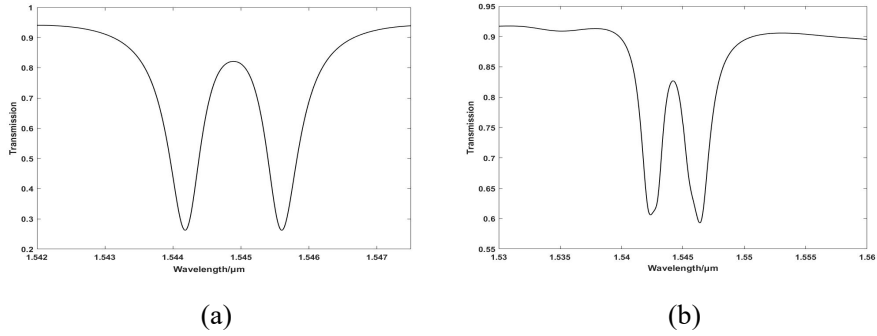

**Fig. 2.** (a) The numerical calculation result, (b) the FDTD simulation result.

In Fig.2(a), the resonance peaks obtained by the numerical calculation is completely symmetrical. In Fig.2(b), the resonance peaks obtained by the simulation is obviously asymmetrical. In fact, the other factors in the design and debug are impossible to be quantified in the theoretical formula, therefore the simulation result relatives to the numerical calculation result has more credibility.

Comparing Fig.2(a) with Fig.2(b), there are a slight fluctuation due to the difference between the wave-guide bending loss and the cladding absorption loss. The wave-guide bending loss decreases with the increasing of the radius. The scattering loss and bending loss are constant when given the MRR cavity size and material.

## 3. COMSOL Analysis

We adopt SU8 wave-guide, in the FDTD simulation software, we use optical field to detect the slot region electric field energy distribution, as shown in Fig.3(a). In the COMSOL simulation software, we use sound field to detect the material deformation, as shown in Fig. 3(b).

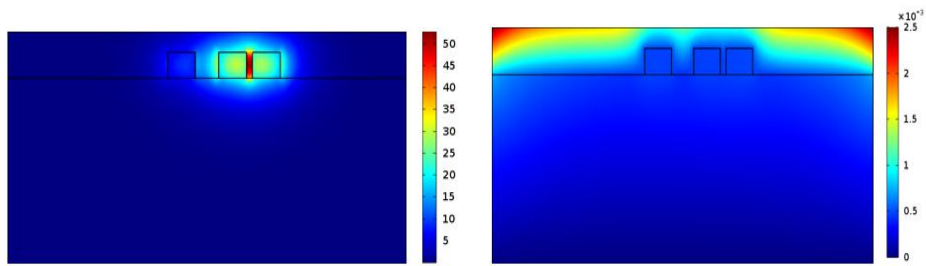

(a) (b)

**Fig.3.** (a)The slot region electric field energy distribution, (b)deformation of the material.

In terms of Fig.3(a), we can see that the electric field energy is concentrated in the slot region.[29] The acoustic pressure distributed around the wave-guide material, which caused by the MRR structure and the properties of material.

In the following, we place ultrasonic sensor into a liquid environment. In order to investigate the MRR structure deformation and sensitivity, we further optimize the optical and acoustic performance of the system. Our system adopts Cantilever arm structure with the receiving surface area $25\mu m \times 60\mu m$, as shown in Fig.4(a). The numerical unit in the legend is micron, the action of external acoustic pressure will lead to hanging end deformation as shown in Fig.4 (b).

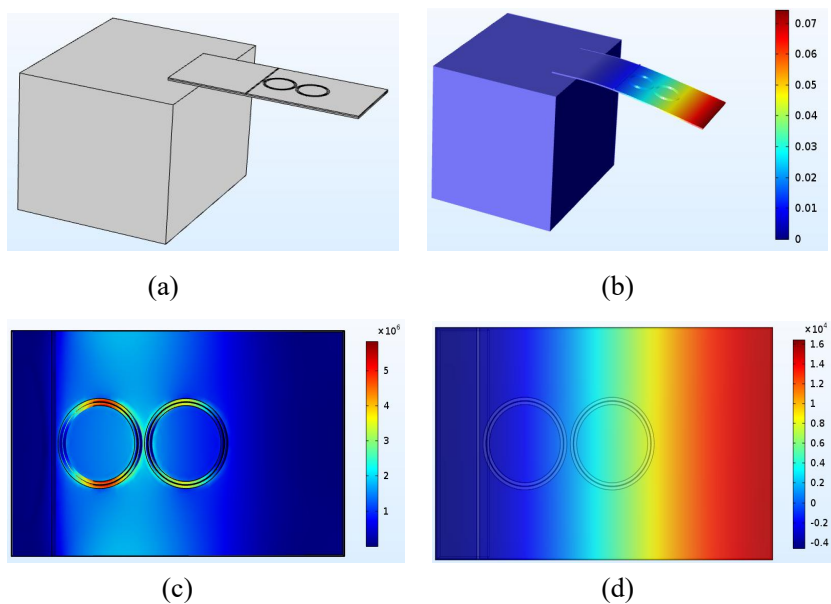

(a) (b)

(c) (d)

**Fig.4.** (a)Sketch of the pressure-induced deformation, (b)the maximum deflection of the membrane is $0.07\mu m$, the pressure above the membrane is $1MPa$, (c)the stress distribution diagram, (d)the sound pressure energy distribution.

Here, the frequency of sound field is $50KHz$, the sound pressure is $1MPa$. We can seen that the maximum position change of deformation is about $0.07\mu m$.

In fact, the detected object usually appears in a different direction. In the following,

we continue to study the change of the $n_{eff}$ of ultrasonic sensor when the sensor receives the sound pressure from different directions and positions.

## 4. The ultrasonic Field Analysis

Assuming the sound field is incident into the sensor, the sound pressure changes the $n_{eff}$ of the ultrasonic sensor, the sensitivity can be calculated according to the rate of change, as shown in Fig.5(a). At the same time, an important problem needs to be paid attention to that the distance between the sound field and the ultrasonic sensor also affects the $n_{eff}$ of the ultrasonic sensor, as shown in Fig.5(b).

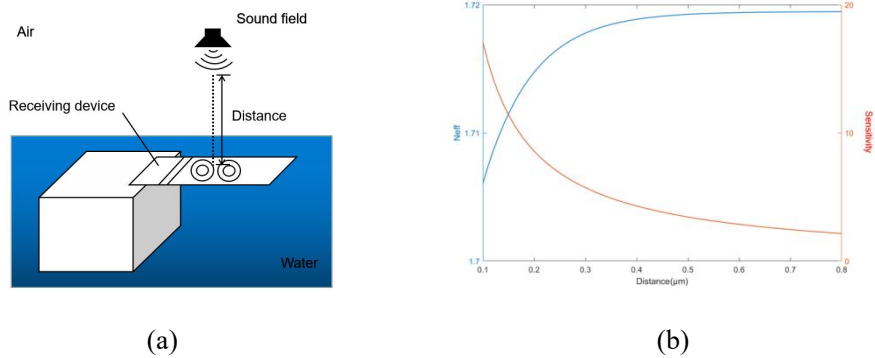

(a)          (b)

**Fig.5.** (a) The position of the sound field, (b) the change in effective refractive index with different position(blue line),the change in sensitivity with different distance (red line).

First, we change the distance between the sound field and the receive plane of ultrasonic sensor, and then analyze the influence of distance on the $n_{eff}$ of the ultrasonic sensor. As shown in Fig.5(a), the influence of distance on the effective refractive index of sensor material is analyzed. As shown in Fig.5(b), the change of the effective refractive index of the material when the distance between the sound field and the sensor is adjusted from $0.1\mu m$ to $0.8\mu m$, we find that the distance has a significant impact on the $n_{eff}$ of the ultrasonic sensor, when the distance is between $0.1\mu m$ and $0.4\mu m$, the influence changes significantly, while the curve gradually flattens when the distance exceeds $0.4 \mu m$. Considering that the distance between the sound field and the ultrasonic sensor is too far, which will lead to the loss of the energy of the sound field, the distance between the sound field and the sensor should be set as $0.5 \mu m$ in this paper. Based on these, the sensitivity curve can be further drawn. In Table.1, we compare the detection accuracy of several devices with different radius.

**Tabel. 1** Comparison of detection range levels of different size devices.

| Hydrophone system | Detection distance/ order of magnitude | Radius/ order of magnitude |
| --- | --- | --- |
| ultrasonic detector[23] | $2 \mu m / 1 \mu m$ | $30 \mu m / 10 \mu m$ |
| Fiber array [30] | $0.1 mm / 0.1 mm$ | $1.25 mm / 1 mm$ |
| Our work | $0.1 \mu m / 0.1 \mu m$ | $5 \mu m / 1 \mu m$ |

In Table.1, we compare the detection accuracy of several groups of devices with different radii. In reference [23], the detector radius is $30 \mu m$, while the detection distance is $2 \mu m$; in reference [30], the detector radius is $1.25 mm$, and the detection distance is $0.1 mm$; in our study, the detector radius is $5 \mu m$, and the detection distance is $0.1 \mu m$. It can be seen that the difference between the detection distance and the radius of the three devices is one order of magnitude, that is, the difference is $10^{-1}$. The following study shows that the detector with the grade difference is more accurate in detecting different distances from the sound field.

Next, we change the direction of the sound field to study the change of the effective refractive index of the sensor material. Fig.6(a) is a model diagram of the sound field. The sound field are received by the sensor in different directions. The angle between the sensor and the sound source is adjusted from -90º to 90º. Fig.6(b) shows the change of effective refractive index corresponds to the change of angle between sensor and sound source.

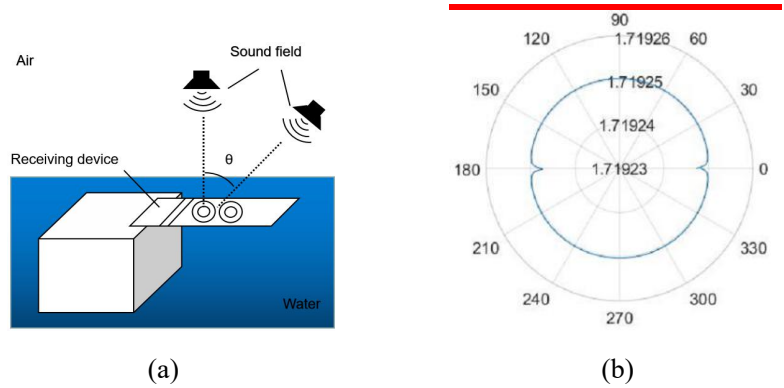

| (a) | (b) |

**Fig. 6.** (a) Sound fields with different angles, (b) the change of effective refractive index with the angle of sound field.

We can see from Fig.6(b) that the change of angle has a little impact on $n_{eff}$ of the acoustic sensor, when the incident angle is zero, the minimum $n_{eff}$ closes to zero. When the change rate of angle becomes bigger, the sensitivity is also becomes higher. Therefore, the change of angle is more sensitive compared with the change of the distance. We usually discuss the situation when the acoustic field is incident into the surface of the sensor(angle is 0° degree).

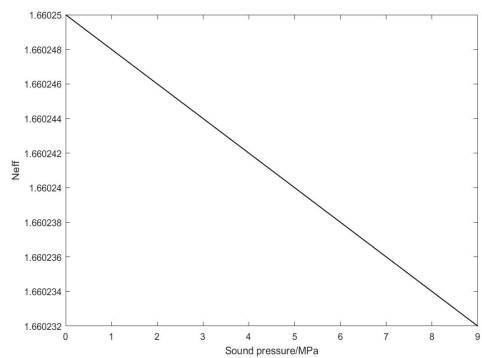

**Fig. 7.** The change of effective refractive index with sound pressure.

The distance between sound field and ultrasonic sensor to be $0.4 \mu m$. The change of the $n_{eff}$ with the increasing of the sound pressure as shown in Fig.7.

In general, the performance of an ultrasound detector is characterized by the sensitivity ($S = \partial T / \partial P$).[33]

$$S = \frac{\partial T}{\partial P} = \frac{\partial T}{\partial \lambda} \frac{\partial \lambda}{\partial n_{eff}} \frac{\partial n_{eff}}{\partial P} = 1462.5 \text{mV/kPa}, \qquad (5)$$

Finally, we complete the performance comparison of our micro-optical system and nano-optical system as ahown in Table 2.

**Table. 2** Comparison of reported hydrophone in different systems.

| Hydrophone system | Frequency | Angle | Distance | wavelength | Radius | Sensitivity |
|---|---|---|---|---|---|---|
| Piezoelectric ceramic[5] | 2.15M$Hz$ | —— | 12.7$cm$ | —— | 47.5$\mu m$ | -31.75dB |
| Fiber array [30] | 3k$Hz$ | -90°~90° | 100$\mu m$ | 1550$nm$ | 1.25$mm$ | -136dB |
| Our work | 50k$Hz$ | -90°~90° | 0.1$\mu m$ | 1545$nm$ | 5$\mu m$ | 1462.5mV/kPa |

From Table 2 we can see that our sound field can detect the smallest object within $0.1\mu m$.

In order to increase the stability of the DMRRs cavity output optical signal, we chose a DMRRs array, which is composed of double DMRRs cavity, as shown in Fig.8(a).

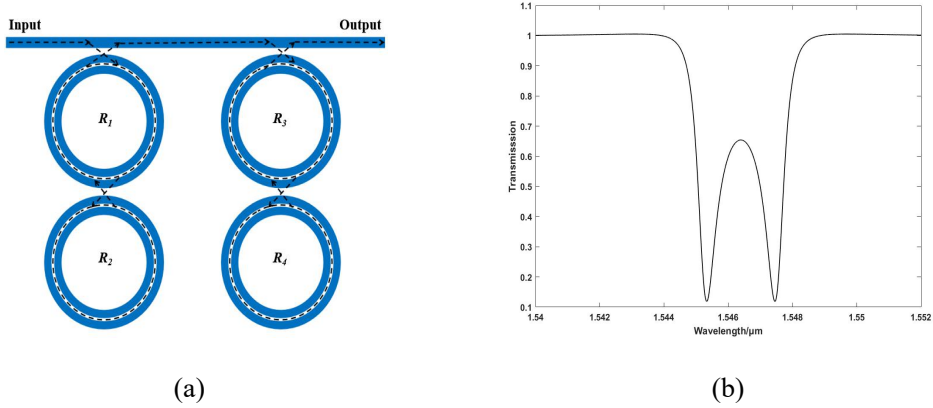

(a)          (b)

**Fig.8.** (a) The structure of DMRRs array, (b)the transmission spectrum.

The transmission spectrum formula can be further deduced by

$$T = 1 - \left| \frac{\delta^2 - 2\delta\mu B_1 - 2\delta t B_2 + 2(\delta + t\mu)B_1 B_2 + \mu^2 t^2 + t^2 B_2^2 - 2\mu B_1^2 B_2 - 2t B_1 B_2^2 + B_1^2 B_2^2}{1 - 2tB_1 - 2\mu B_2 + 2(\delta + t\mu)B_1 B_2 + t^2 B_1^2 + \mu B_2^2 - 2t\delta B_1^2 B_2 - 2\mu\delta B_1 B_2^2 + \delta^2 B_1^2 B_2^2} \right|^2, \qquad (6)$$

where $\delta = (\xi^2 + \eta^2)t/2$ and $\xi = 0.943$ is the coupling coefficient between the straight wave-guide and the ring, $\eta = 0.508$ is the coupling coefficient between two different rings and $t=0.75$ is the transfer coefficient between wave-guides. $B_1=\alpha e^{j\theta}$, $B_2=\alpha e^{-j\theta}$, $\alpha=0.9$ is the transmission loss of the ring and $\theta=2\pi L/\lambda$ is the phase shift, where $L$ is the circumference of a ring and $\lambda$ is the wavelength of light.

Comparing Fig.8(b) with Fig.2(a), we find that optical spectrum of the double DMRRs array is obviously better than that of the single DMRRs. Therefore, the multi-DMRRs array can significantly improve the optical properties of DMRRs system.

It can be seen from the Table 3 that our system needs the minimum radius can obtain the corresponding optical and acoustic performance. We use the Cantilever arm structure to complete the detection when the sound source with low frequency.

Table 3. Comparison of reported MRR hydrophone.

| MRR hydrophone | NEP (Pa) | Feq (MHz) | $\lambda$ (nm) | R ($\mu m$) | S (mV/kPa) | Ring | Q-factor |
|---|---|---|---|---|---|---|---|
| polymer micro-ring[20] | 1M | 50 | 1550 | 47.5 | 21pm/MPa | one | $6\times10^2$ |
| polymer micro-ring[21] | 21.4 | 20 | 769 | 25 | 66.7 | one | $4\times10^5$ |
| ultrasonic detector[23] | 105 | 350 | 772 | 25 | 20.05 | one | $1.3\times10^5$ |
| slot micro-ring[28] | 1M | 540 | 1550 | 12 | 66.7 | one | $8.34\times10^8$ |
| slot micro-ring[29] | —— | 150 | 1372 | 5 | 2453.7 | two | $1.24\times10^6$ |
| our work | 1M | 0.08 | 1545 | 5 | 1462.5 | two | $1.54\times10^3$ |

It can be seen from Table 3 that the micro-ring resonator of our system has the optimized optical and acoustic properties with the smallest radius.

## 5. Conclusion

When the light field is incident into the DMRRs and the sound pressure propagates along the surface, both the stress deformation and elastic-optical effect have some affect on the $n_{eff}$ of the DMRRs. These further influence the output spectrum of acoustic sensor. The Cantilever arm structure can more obviously shows the stress deformation of the slot-based DMRRs. We analyze the system's response to the sound field incident from different angles and the change of the sensor's $n_{eff}$ with the distance between the sound field and the sensor. The $n_{eff}$ of the acoustic sensor rises up with the increasing of sound field position. The volume of our acoustics sensor is much smaller than the traditional acoustics sensor. We can achieve precise positioning of object within $0.1 \mu m$. Finally, we verify that the DMRRs array can provides a more stable output spectrum. Therefore, our acoustic sensor can be used as a hydrophone when it comes to underwater detecting.

**Acknowledgements**

This work was supported by National Natural Science Foundation of China [grant number 11504074] and the State Key Laboratory of Quantum Optics and Quantum Optics Devices, Shan xi University, Shan xi, China [grant number KF201801].

**References**

[1] Yanagihara N, Gyo K, Suzuki K 1983 *Ann. Otol., Rhinol., Laryngol* 92 223

[2] Felix N, Tran-Huu-Hue L P, Walker L 2000 *Ultrasonics* 38 127

[3] Zhang G J, Xu Q D, Zhang C 2020 *Sensors and Actuators A:Physical* 306 111969

[4] Yang X B, Yin G J, Tian Y 2019 *IEEE Transactions on Ultrasonics, Ferroelectrics, and Frequency Control* 67 356

[5] Wang J S, Xiao J J, Zhao X Y 2020 *Ceramics International* 46 11913

[6] Scott F, Alexei T, Mark M 2005 *17th International Conference on Optical Fibre

*Sensors* 5855 627

[7] Hill D J, Hodder B, Freitas J D 2005 *17th International Conference on Optical Fibre Sensors* 5855 904

[8] Xu S Y, Huang H, Cai W C 2019 *Appl. Opt.* 58 7774

[9] Li H, Sun Q Z, Liu T 2019 *J. Lightwave Technol.* 38 929

[10] Saijyou K, Okuyama T, Nakajima Y 2016 *IEEE J. Oceanic Eng.* 41 373

[11] Marcatili E A J 1969 *Bell Syst. Tech. J* 48 2103

[12] Scheuer J, Paloczi G T, Yariv A 2005 *Appl. Phys. Lett.* 87 251102

[13] Van Camp M A, Assefa S, Gill D M 2012 *Opt. Exp.* 20 28009

[14] Hassan A U, Hodaei H, Miri M A 2015 *Phys. Rev. A* 92 063807

[15] Zhao C Y, Zhang L, Zhang C M 2018 *Opt. Commun.* 414 212

[16] Karimi F, Knezevic I 2019 *Opt. Materials Exp.* 9 4456

[17] Serna S, Lin H T, Alonso-Ramos C, Lafforgue C, Xavier L R, Richardson K, Cassan E, Dubreuil N, Hu J J, Vivien L 2019 *Opt. Lett.* 44 5009

[18] Dell'olio F, Passaro V M N 2007 *Opt. Exp.* 15 4977

[19] White I M, Fan X 2008 *Opt. Exp.* 16 1020

[20] Chao C, Ashkenazi S, Huang S, O'Donnell M, Guo L J 2007 *IEEE Trans. Ultrason. Eng.* 54 957

[21] Ling T, Chen S, Guo L J 2011 *Appl. Phys. Lett.* 98 204103

[22] Rekha R, Sudarshan S M, Shushma M R 2011 *TENCON 2011-2011 IEEE Region 10 Conference* 21 747

[23] Zhang C, Ling T, Chen S, Guo L J 2014 *ACS Photonics* 1 1093

[24] Almeida V R, Xu Q F, Barrios C A, Lipson M 2004 *Opt. Lett.* 29 1209

[25] Barrios C A, Gylfason K B, Sanche B 2007 *Opt. Lett.* 32 3080

[26] Barrios C A, Banuls M J, Gonzalez-Pedro V 2008 *Opt. Lett.* 33 708

[27] Robinson J T, Chen L, Lipson M 2008 *Opt. Exp.* 16 4296

[28] Zhang S, Chen J, He S 2017 *Opt. Commun.* 382 113

[29] Zhang C M, Zhao C Y 2019 *Optik* 178 1029

[30] Wang J Y, Ai F 2018 *Opt. Exp.* 26 25293

[31] Chao C Y, Ashkenazi S, Huang S W 2007 *IEEE Transactions on Ultrasonics, Ferroelectrics, and Frequency Control* 54 957

[32] Zhao C Y, Chen P Y, Li P Y, Zhang C M 2020 *Int. J. Mod. Phys. B* 34 2050145

[33] Li H, Dong B Q, Zhang Z 2014 *Sci. Rep.* 4 4496